\documentclass[aps,prl,showpacs,amsmath,nofootinbib
,superscriptaddress,twocolumn,preprintnumbers]{revtex4}

\usepackage{natbib}
\usepackage{amsmath}
\usepackage{amssymb}
\usepackage{graphicx}
\usepackage{color}

\begin{document}

\setlength{\unitlength}{1mm}
\title{Temporal Variation of the Fundamental Physical Quantities in a Static Universe}
\author{Meir Shimon}
\affiliation{School of Physics and Astronomy, Tel Aviv 
University, Tel Aviv
69978, Israel}
\email{meirs@wise.tau.ac.il}

\begin{abstract}
The standard interpretation of the observed 
redshifted spectra and luminosities towards distant 
astrophysical objects is that the universe is expanding, 
an inference which is found to be consistent with other 
cosmological probes as well. Clearly, only the 
interpretation of {\it dimensionless} quantities does 
not depend on the physical unit system as opposed to 
{\it dimensional} quantities whose dynamics does 
depend on the arbitrarily chosen 
system of units. Indeed, cosmological redshift is a 
{\it dimensionless} quantity (related to the {\it relative} 
expansion between cosmological and local length scales) 
but its conventional interpretation as an evidence for 
universe expansion is based on dimensional quantities; 
it results from selecting local length scales as standard 
rulers, i.e. fixed constants. All that redshift or 
luminosity measurements really indicate is that 
cosmological scales expand {\it relative} to 
local scales.
An alternative choice of a ruler could be the 
distance between two remote galaxies, in which 
case local distances have to decrease with 
time (with respect to the ruler) for consistency with 
redshift measurements. In the latter choice, 
microscopic scales such as the 
Compton wavelength, or the Planck length, decrease with 
time, and consequently fundamental `constants' such as 
the Planck constant, speed of light, Newton gravitational 
constant, and particle masses, are spacetime-dependent. 
To illustrate this fundamental indeterminacy we 
construct an alternative interpretation to the expanding 
model that is characterized by a static metric 
with time-dependent fundamental physical quantities. 
The two alternative descriptions, are referred to as the 
`expanding' and `static' space perspectives, respectively.
Cosmological inflation, recombination, and all other early 
universe processes are unaltered; the `expanding' 
and `static' perspectives are associated with exactly the 
same cosmological model.
One could equally well choose the ruler from a continuum of scales 
spanned by these extremes; cosmological and local scales. 
For each such a choice, cosmological and local scales expand 
and contract, respectively, but with slower rates than in the 
`expanding' and `static' perspectives, resulting in a continuum 
of observationally indistinguishable interpretations.
\end{abstract}


\maketitle

{\bf Introduction}.-- The early days of modern cosmology 
saw a few revolutionary theoretical and observational feats 
that shaped our understanding of the universe on cosmological 
scales. Since the inception of general relativity (GR) 
in 1916 our notions of dynamical spacetime have been 
fundamentally revolutionized. In 1929 Edwin Hubble 
formulated his famous law [1], and in 1920's Friedmann, 
Lemaitre, Robertson, and Walker (FLRW) put forward an 
expanding universe model; a solution to Einstein field 
equations describing a homogeneous and isotropic background 
metric [2,3]. Other popular models have been proposed, e.g. 
the steady state universe by Hoyle, Bondi and others [4,5].
A long debate over the cosmological model has virtually 
ended in 1965, with the discovery of the cosmic microwave 
background (CMB) radiation by Penzias and Wilson [6]; 
explaining the thermal nature of the CMB within frameworks 
other than the big bang model seem largely unnatural, 
essentially ruling out big bang's rival theories [7]. 

A few seemingly puzzling global features of the universe 
were elegantly explained at the early 1980's by a single 
theory -- cosmological inflation -- that invokes a 
hypothetical scalar field (or fields), driving space into 
a very brief but significant exponential expansion at 
the very beginning of the universe. Within this framework, 
the observed spatial flatness of the universe, 
underabundance of magnetic monopoles, etc., are turned 
from nuisances to natural predictions [8, 9]. 

The spectrum of phenomena explained by the standard 
cosmological model spans scales from the largest down 
to microphysical processes that took place in the first 
few minutes after the big bang, e.g., the electroweak 
phase transition, the formation of light nuclei, i.e. 
big bang nucleosynthesis (BBN), as well as 
recombination, structure formation, etc., at latter 
times. According to the standard picture, these 
processes are controlled by space expansion, which 
provides the `clock' for the cooling down of matter 
and radiation. The elegance of this description of 
the universe with only a dozen free parameters 
cannot be overstated. 

Aside from minor modifications that may be required, 
the big bang paradigm provides an extremely successful 
description of a multitude of cosmological phenomena, 
incorporating very remote disciplines and uniting 
micro- and macrophysics. Yet, is it the only possible 
{\it interpretation} of the observable universe ? 
{\it The notion of space expansion 
clearly depends on the system of units and therefore 
does not constitute a unique interpretation of 
the observed dimensionless cosmological redshift.}
In this {\it letter} an alternative perspective is proposed  
featuring a flat, non-expanding, metric but the physical 
constants are implicitly time-dependent. The proposed 
`static space perspective' (SSP) is observationally 
indistinguishable from the conventional FLRW 
interpretation, which is referred to in this work as 
the `expanding space perspective' (ESP). In a companion 
letter [10] it is shown that gravitation can be viewed as 
a prescription for a local {\it redefinition} of the 
standard rulers of physical units. In this {\it letter}, 
the idea is applied to the cosmological case.
This does not conflict with upper bounds on 
spacetime variations of the physical constants because 
according to the SSP all dimensionless physical quantities 
are exactly the same as in the ESP; observers simply 
cannot distinguish their meter contraction from 
space expansion. The ESP and SSP are just the two 
extremes of a continuum of possible interpretations 
of the dynamics of the universe; this non-uniqueness 
results from focusing on the dynamics of dimensional 
quantities.

The basic idea underlying the SSP is described in 
the next section, followed by a derivation of the 
scaling transformations of fundamental constants 
from the ESP to SSP in section 3. The results 
are summarized and discussed in section 4. 

{\bf The Emergence of Cosmological Expansion}.-- 
A fact underlying the SSP is that the action for a 
point particle of mass $m$ is 
\begin{eqnarray}
S=\int mc\cdot ds=\int mc\cdot \frac{ds}{d\lambda}d\lambda
\end{eqnarray}
where the infinitesimal interval is defined as
\begin{eqnarray}
ds^{2}=g_{\mu\nu}dx^{\mu}dx^{\nu}
\end{eqnarray}
and Greek indices run over all four spacetime dimensions, 
summation convention implied, $\lambda$ is an 
affine parameter, and $g_{\mu\nu}$ is the metric. 

Cosmology is usually thought of as 
a purely classical field, and indeed it is. However, 
tight observational constraints on the variation 
of the dimensionless fine structure constant 
$\alpha=e^{2}/(\hbar c)$ require that any variation 
of either $c$ or $e$ is compensated by variation 
of $\hbar$. Here, quantum mechanics enters the 
formulation simply because emission mechanisms, which 
are integral part of the cosmological paradigm, are 
quantum mechanical in nature. 
In addition, energy quanta of photons are $\hbar\nu$; thus, one 
is compelled to incorporate any spacetime dependence of $\hbar$ in 
the formulation of a consistent cosmological model. 
The relation between dimensionless quantities and quantum 
mechanics is further highlighted in [10].
The transition amplitude for a particle to propagate from a 
quantum state $\psi_{1}$ to $\psi_{2}$ is given 
by $\Gamma_{1\rightarrow 2}=\sum_{hist}\exp(iS/\hbar)$ where the 
sum runs over all possible particle histories 
(trajectories) subject to fixed initial and final states, 
$\psi_{1}$ and $\psi_{2}$, respectively [11]. 
If $mc/\hbar$ is spacetime-dependent then minimizing the
phase $\phi\equiv \int dS/\hbar$ is akin to minimizing 
an interval with an effective metric 
$\tilde{g}_{\mu\nu}\propto (mc/\hbar)^{2}g_{\mu\nu}$.
This would consequently modify the geodesic equations 
and induce an {\it apparent} cosmological redshift, as 
will be discussed below. 

Let us start with the case of a massive free point particle 
in flat spacetime and comment on the massless particle 
case later. The infinitesimal interval is
\begin{eqnarray}
ds^{2}=\eta_{\mu\nu}dx^{\mu}dx^{\nu}
\end{eqnarray}
where $\eta_{\mu\nu}=diag(1,-1,-1,-1)$ is the Minkowski 
metric. The minimization of $\phi$ consists effectively 
of finding the geodesics of a curved spacetime described 
by the metric 
$\tilde{g}_{\mu\nu}\propto (mc/\hbar)^{2}\eta_{\mu\nu}$. 
Defining
\begin{eqnarray}
a(t,{\bf x})\equiv\lambda_{c}(t_{0},{\bf x}_{0})/
\lambda_{c}(t,{\bf x}),
\end{eqnarray}
i.e., the ratio of the Compton wavelength, 
$\lambda_{c}\equiv\hbar/(mc)$, here and now to its value at 
$(t ,{\bf x})$, and further assuming that $a$ is a function 
of time only, one arrives at the FLRW metric written in 
the conformal gauge
\begin{eqnarray}
ds^{2}=a^{2}(t)\eta_{\mu\nu}dx^{\mu}dx^{\nu}
\end{eqnarray}
if $a$ is identified with the `scale factor'.
Here, $x^{\mu}=(c\tau,x,y,z)$ where $d\tau\equiv dt/a(t)$ 
with $\tau$ and $t$ the conformal, and Newtonian cosmic 
times, respectively. Breaking from the standard procedure, 
the speed of light is redefined, $c\rightarrow c/a$, while 
the time coordinate, $t$, is held unchanged. As the time 
coordinate is unchanged time scales for physical processes 
should be exactly the same in both the ESP and SSP. 
This condition will be supplemented later by another 
constraint that completes the definition of the SSP.
Alternatively, one could define a different universal 
scaling for all time scales while retaining consistency 
with observations, which merely amounts to a redefinition 
of the time unit [10].

Let us recall Einstein field equations for a FLRW metric 
in the Newtonian gauge for reference [12-15]
\begin{eqnarray}
\left(\frac{\dot{a}}{a}\right)^{2}
+\frac{k}{a^{2}}&=&\frac{8\pi G}{3c^{2}}\rho
+\frac{\Lambda c^{2}}{3}\nonumber\\
2\frac{\ddot{a}}{a}
+\left(\frac{\dot{a}}{a}\right)^{2}+\frac{k}{a^{2}}&=&
-\frac{8\pi G}{c^{2}}P+\Lambda c^{2}
\end{eqnarray}
where $\rho$, $P$, and $\Lambda$ are the energy density, 
pressure, and cosmological constant, respectively, and 
overdot represents the time derivative. For simplicity, 
we consider the case of a spatially flat universe ($k=0$). 
In the conventional 
ESP, the energy density and pressure on the right hand 
side determine the dynamics of the expansion scale factor 
$a(t)$. The analog statement in the SSP 
would be that the energy and pressure content of the 
universe determine the time evolution of the Compton 
wave numbers and since classically 
the metric is obtained by integrating Einstein equations, 
i.e. by performing double integration of
the energy momentum source term over time, 
then particles/fields local characteristics such 
as mass, electric charge, etc. (as will be shown below), 
are all determined by the energy-momentum content of 
the entire universe. This interpretation
is very surprising 
and seemingly counter intuitive simply because 
of our prejudice 
and our standard physical units. In the following we elaborate on this 
and show that the SSP provides a very 
consistent picture of the universe, exactly as 
the ESP does.

Note that our starting point, Eq.(1), is valid for 
massive particles only, although historically the first 
observational indication for space expansion came from 
redshifted radiation. One can replace Eq.(1) with [16]
\begin{eqnarray}
S=\frac{1}{2}\int d\lambda\left[\chi^{-1}
g_{\mu\nu}\dot{x}^{\mu}\dot{x}^{\nu}-\chi(mc)^{2}\right]
\end{eqnarray}
where $\chi$ is an auxiliary field. 
Eq.(7) applies to the massless particle case as well.
It can be readily verified that if the field $\chi$ 
is ascribed with units of inverse mass the exact same 
picture of emergent space expansion 
from static flat metric is obtained.

{\bf Fixing the Scaling of Absolute Physical Units}.--
Any dimensionless quantity is 
independent of our choice of the physical 
system of units, and therefore for the ESP and SSP 
to be observationally indistinguishable it is 
sufficient to require that all dimensional quantities 
of the same physical dimension transform similarly 
from the ESP to SSP [10]. This 
`scaling universality', together with 
$\lambda_{C}\propto a^{-1}$ (Eq. 4), implies 
that all length scales are $\propto a^{-1}$ 
(in ESP terminology they are said to be `comoving').
Since our construction requires that time scales in 
SSP and ESP agree, the rest energy of a massive 
particle $E_{m}=mc^{2}$ must scale as $\hbar$. 
Combining these two relations one obtains that the 
speed of light is varying over cosmological times 
(see discussion below Eq. 5)
\begin{eqnarray}
c=c_{0}/a(t)
\end{eqnarray}
where $a$ is a solution of the FLRW equations (Eq. 6). 
Here, $a$ is interpreted as the time variation of $c$ 
and other `constants', not as the expansion 
scale factor.

In the SSP the photon wavelength does not stretch 
but since the speed of light is a monotonically 
decreasing function of time then the frequency 
redshifts
\begin{eqnarray}
\nu=c/\lambda=\nu_{0}/a
\end{eqnarray} 
which explains the observed redshift from the SSP in a 
completely different manner. Observing 
spectra of astrophysical objects involves detecting 
flux of light quanta at some energy bands. 
These energy bands are then conventionally converted 
to wavelengths assuming that $c$ and $\hbar$ are 
known. From the SSP, shifting features on a spectrum 
diagram may be simply due to the variation of $c$ and 
$\hbar$ between light emission and reception. 

The energy density of pressureless species, $\rho_{m}$, 
scales as $a^{-3}$ from the ESP (due to space expansion).
A second arbitrarily chosen condition that supplements 
the invariance of time units (see below Eq. 5), is the 
invariance of energy density under transformation of units 
from the ESP to SSP.
Since space is static in the SSP and $c\propto a^{-1}$, 
then the rest energy $E_{m}=mc^{2}$ must scale as 
$a^{-3}$, and therefore,
\begin{eqnarray}
m=m_{0}/a.
\end{eqnarray}
Since $E$ scales as $h$ ($h=2\pi\hbar$) 
in the SSP one must conclude that
\begin{eqnarray}
h=h_{0}/a^{3}.
\end{eqnarray}
Combined with Eq.(9) and the fact that space is static 
in SSP one obtains $h\nu\propto a^{-4}$, i.e. the energy 
(and therefore also the energy density, $\rho_{\gamma}$) 
of radiation scales as $a^{-4}$, as 
desired. This construction recovers the standard 
result familiar from the ESP that the energy density 
scales as $a^{-3}$ and $a^{-4}$ in the dust 
and radiation cases, respectively [12-15]. Carrying this 
relation over from the ESP to SSP is crucial to the 
theory of structure formation and neutrino free 
streaming [12-15].

Next, the scaling of Newton gravitational constant 
is determined. In the Newtonian limit the gravitational 
acceleration experienced by a point particle a 
distance $r$ from a massive object with mass $M$ is
\begin{eqnarray}
\ddot{r}=-GM/r^{2}.
\end{eqnarray}
Since lengths and masses scale as $a^{-1}$ (Eqs. 4 \& 10), 
for time scales not to vary between the two perspectives, 
Newton's gravitational constant must scale as
\begin{eqnarray}
G=G_{0}/a^{2}.
\end{eqnarray}  
To determine the electric charge scaling note that for 
the charged particle potential energy $e^{2}/r$ to scale 
as $a^{-3}$ the electron charge scaling must be
\begin{eqnarray}
e=e_{0}/a^{2}.
\end{eqnarray} 
Alternatively, for the dimensionless fine structure 
constant, $\alpha=e^{2}/(\hbar c)$, to be the 
same in the ESP and SSP, $e=e_{0}/a^{2}$. 
Therefore,  the (dimensionless) 
relative strength of the electric and gravitational 
forces $F_{e}/F_{g}=e^{2}/(GmM)$ is independent of 
time, in contrast to Dirac's vision of unifying the 
electric and gravitational forces at the remote past 
by assuming that $G$ is monotonically decreasing 
with time [17]. Clearly, the construction of the 
SSP simply amounts to a time-dependent redefinition 
of the physical `constants' and involves no new physics.

The Rydberg constant $R_{\infty}=m_{e}c\alpha^{2}/(2h)$ 
that has inverse length units controls the wavelengths 
of emitted radiation in atomic transitions. It is 
trivial to verify that it properly transforms from the 
ESP to SSP, thereby mimicking cosmological redshift 
(compare to Eq. 4). 
From the SSP, the observed cosmological redshift is 
simply due to variation of atomic `constants' over 
cosmological time while space remains static. 

The expanding space `clock' of the ESP 
is replaced with the intrinsic time dependence of the 
physical `constants' in the SSP.
This is nicely illustrated 
with cosmological recombination. 
As time scales in the ESP and SSP are 
identical it is clear that the SSP explains the 
thermal CMB exactly as ESP does. From the SSP the 
primordial electron-photon plasma was tightly coupled 
not because of extremely high number densities. Rather, 
it is the diverging electron charge and speed of light 
at earlier times which are responsible for the tight 
electron-photon coupling and ensuing thermal 
equilibrium of the CMB. Recombination, from this 
perspective, took place when $a$ was sufficiently large 
to suppress $e$ 
and $c$ and stretch the photon mean free path, 
$\lambda_{f}=c(t)t_{C}$, to horizon scales or 
larger. Note that the Compton time, 
$t_{C}^{-1}=n_{e}\sigma_{T}c$, trivially transforms, 
by construction. More specifically, according to the 
ESP it scales as $\propto a^{-3}$ because the 
electron number density falls off with the expanding 
universe. In contrast, from the SSP $n_{e}$ is 
constant but $c$ and $\sigma_{T}$ fall off as 
$a^{-1}$ and $a^{-2}$, respectively, again 
resulting in an overall $a^{-3}$ scaling. 

Note that, consistent with the requirements of 
constancy of time scales and variation of length 
and mass scales, the fundamental Planck unit system 
scales as $t_{P}=\sqrt{\hbar G/c^{5}}\propto const.$, 
$l_{P}=\sqrt{\hbar G/c^{3}}\propto a^{-1}$, 
and $M_{P}=\sqrt{\hbar c/G}\propto a^{-1}$, as can 
be easily verified using Eqs.(8)-(13).

As briefly discussed above, it is required that the 
history and details of structure formation remain 
unchanged when allowing the physical `constants' to vary 
with time. Indeed, from the transformation of $G$, 
$c$, $\rho$, and $\Lambda c^{2}$ (recall that the 
latter has units of $time^{-2}$ and is therefore 
invariant) between the ESP and SSP, 
and the fact that energy density and pressure 
are the same in both perspectives [10], it is 
clear that the FLRW equations, Eqs. (6), as well as 
their linearly perturbed versions, are exactly the 
same in both perspectives. Small perturbations 
of the FLRW metric and the energy-momentum 
tensor are responsible for structure formation and 
therefore this last observation completes the 
argument that structure formation is unchanged 
when viewed from either perspective. From a more 
fundamental viewpoint, the ESP and SSP are only 
two out of infinitely many possible representations 
of the same physical reality since the 
quantum phases associated with both descriptions 
are identical [10].

{\bf Summary}.-- The expanding universe scenario 
is probably one of the most impressive achievements 
of modern physics. With only limited access to high 
energies ($\lesssim TeV$) our theories of particle 
physics, principles of quantum mechanics, and GR and 
theories of structure formation, have been combined 
over the last few decades into a coherent picture of 
the universe on cosmological scales and back to its 
very beginning. 

In the conventional 
system of units, values of masses, Compton wavelengths, and angular 
momentum quanta ($\hbar$) of elementary particles are fixed, but not others, 
perhaps because of the way physics historically evolved from measurements 
of processes involving elementary particles and only later with 
observations of the `expanding' universe. Had it evolved otherwise, it is 
not unlikely that cosmological distances would have been considered fixed 
(i.e. selected as rulers) 
at the expense of decreasing Compton wavelength with time. The standard 
lore could have then been that space is static but the fundamental 
microscopic quantities are decreasing with time.    

Here, an alternative interpretation to the expanding 
universe picture is proposed. 
Motivated by the homogeneous and isotropic FLRW 
description of the universe in comoving frame it is 
proposed that the metric is static at the cost of 
allowing the physical `constants' to implicitly 
depend on time, via the `scale factor' $a$ (here 
adopting the ESP terminology but with a different 
meaning; scale factor of the fundamental 
time-dependent length ruler, not of expanding space).
The resulting picture is fundamentally different, 
yet intimately related to the standard cosmological 
narrative. 

Phenomena such as cosmological redshift, energy 
density evolution, recombination, etc., are explained 
from the SSP in a completely different fashion. These 
two descriptions are shown to yield 
equivalent predictions and are therefore 
observationally indistinguishable. This is expected 
since the transformation from ESP to SSP is equivalent 
to a certain spacetime-dependent squeezing of the 
fundamental rulers [10]. 

Cosmological inflation, according to SSP, is an 
incredibly fast contraction of the Compton 
wavelength of particles. The spatial flatness is 
explained by the immensely decreasing speed of 
light; it implies that only a small fraction of 
the universe is actually observed,
effectively zooming in on a tiny 
fraction of the universe which must therefore be 
nearly spatially flat. Similarly, the horizon 
problem is explained away by the exponential 
deflation of the speed of light; from the SSP, 
in a static universe seemingly causally 
disconnected patches of the universe have been 
in causal contact in the remote past when the 
speed of light was many orders of magnitudes 
larger. The observational absence of magnetic 
monopoles from the ESP is explained by the 
exponential expansion of space during inflation 
where virtually all monopoles have been pushed 
beyond the horizon by the end of inflation. From 
the SSP it is the exponential suppression of the 
speed of light, and thereby of the observable 
fraction of the universe, that is responsible 
to the fact that 
the observable universe is many orders of 
magnitudes smaller than it could be if not for 
`cosmological deflation'. In our `small' universe 
the detection of even a single magnetic monopole 
is very unlikely.

The two viewpoints loose their predictive power at 
a singularity $\approx$14 Gyrs ago; the spacetime 
curvature singularity in the ESP at $a=0$ 
is replaced with divergence of physical constants 
at $a=0$ in the SSP. Although 
the scalar curvature is, by definition, invariant 
to coordinate transformations it is known not to be 
invariant under the action of conformal 
transformations, which is essentially the transformation 
from the ESP to the SSP; this allows replacing the divergent 
FLRW at the big bang (ESP) with a non-singular static 
flat metric (SSP). Indeed, the dimensionless 
curvature should be unaltered under changes of the 
unit system and for this end the notion of curvature 
scalar has to be generalized [10].
The duality between the ESP and SSP is in many 
respects tautological. The {\it relative} expansion 
of the (cosmological) photon and (local) 
Compton wavelength is the common 
key feature to these conceptually radically different 
perspectives. It is interesting to note that the same 
observational reality can be explained by two nearly 
orthogonal perspectives, as well as by infinitely 
many others not discussed here. This should not be 
surprising since the fundamental theory is formulated 
in terms of {\it dimensional} quantities while, 
obviously, the interpretation of the dynamics of 
only {\it dimensionless} quantities is unambiguous [10].

As has been repeatedly argued here 
and elsewhere in the literature, the notion of 
varying constants is only meaningful when these 
constants are dimensionless, but what 
are those `constants of physics'? 
Is the length ruler the Compton wavelength of the 
electron, or the distance between two distant 
galaxies? or some scale in between ? 
Clearly, it is a matter of arbitrary choice, or 
possibly the way physics historically evolved. 
This implies that dynamics must apply 
only to dimensionless quantities whether we 
consider them as `constants' (e.g. $\alpha$) 
or not. In particular, since both the ESP and SSP 
are described in terms of dimensional quantities,  
neither one provides an unambiguous interpretation 
of cosmological observations. 
From that perspective, the notion 
of expanding space or contracting Compton wavelength 
should be replaced with, e.g. expanding dimensionless 
volume. The latter is an unambiguous statement, independent 
of the system of units, and in fact this is what is 
being observed. Putting forward SSP was 
primarily intended to illustrate that dimensional 
formulation of physical theories necessarily leads to 
ambiguous interpretations [10], obviously a well-known fact but 
to the same extent largely ignored in cosmological text books 
and in virtually all scientific work in the field.

\end{document}